 \documentclass[final,5p,times,twocolumn]{elsarticle}

\usepackage{amssymb}
\usepackage{amsthm}
\usepackage{amsmath}

\usepackage[T1]{fontenc}
\journal{Nuclear Instruments and Methods in Physics Research A}

\begin{document}

\begin{frontmatter}



\title{The CYGNO experiment, a directional detector for direct Dark Matter searches}


\author[inst1]{Fernando	Domingues Amaro} 
\affiliation[inst1]{organization={LIBPhys; Department of Physics; University of Coimbra; 3004-516 Coimbra; Portugal}%Department and Organization
}

\author[inst2,inst3]{Elisabetta Baracchini} 
\affiliation[inst2]{organization={Gran Sasso Science Institute; 67100; L'Aquila; Italy}%Department and Organization
}
\affiliation[inst3]{organization={Istituto Nazionale di Fisica Nucleare; Laboratori Nazionali del Gran Sasso; 67100; Assergi; Italy}%Department and Organization
}

\author[inst4]{Luigi Benussi} 
\author[inst4]{Stefano Bianco}
\author[inst4]{Cesidio Capoccia}
\author[inst4,inst5]{Michele Caponero} 
\affiliation[inst4]{organization={Istituto Nazionale di Fisica Nucleare; Laboratori Nazionali di Frascati; 00044; Frascati; Italy}%Department and Organization
}
\affiliation[inst5]{organization={ENEA Centro Ricerche Frascati; 00044; Frascati; Italy}%Department and Organization
}

\author[inst6]{Danilo Santos Cardoso} 
\affiliation[inst6]{organization={Centro Brasileiro de Pesquisas F\'isicas; Rio de Janeiro 22290-180; RJ; Brazil}%Department and Organization
}

\author[inst7]{Gianluca Cavoto} 
\affiliation[inst7]{organization={Dipartimento di Fisica; Universit\`a La Sapienza di Roma; 00185; Roma; Italy}%Department and Organization
}
\affiliation[inst8]{organization={Istituto Nazionale di Fisica Nucleare; Sezione di Roma; 00185; Rome; Italy}%Department and Organization
}

\author[inst2,inst3]{Andr\'e Cortez} 

\author[inst9,inst12]{Igor Abritta Costa} 
\affiliation[inst9]{organization={Dipartimento di Matematica e Fisica; Universit\`a Roma TRE; 00146; Roma; Italy}%Department and Organization
}
\affiliation[inst12]{organization={Istituto Nazionale di Fisica Nucleare; Sezione di Roma Tre; 00146; Rome; Italy}%Department and Organization
}

\author[inst4]{Emiliano Dan\'e}
\author[inst2,inst3]{\underline{Giorgio Dho}}
\author[inst2,inst3]{Flaminia Di Giambattista}

\author[inst8]{Emanuele Di Marco}
\author[inst8]{Giulia D'Imperio}
\author[inst8]{Francesco Iacoangeli}
\author[inst6]{Herman Pessoa Lima J\`unior}

\author[inst10]{Guilherme	Sebastiao Pinheiro	Lopes}
\affiliation[inst10]{organization={Universidade Federal de Juiz de Fora; Faculdade de Engenharia; 36036-900; Juiz de Fora; MG; Brasil}%Department and Organization
}

\author[inst4]{Giovanni Maccarrone}
\author[inst1]{Rui Daniel Passos Mano}

\author[inst11]{Robert Renz Marcelo Gregorio}
\affiliation[inst11]{organization={Department of Physics and Astronomy; University of Sheffield; Sheffield; S3 7RH; UK}
}

\author[inst2,inst3]{David Jos\'e Gaspar Marques}
\author[inst4]{Giovanni Mazzitelli}
\author[inst11]{Alasdair	Gregor	McLean}
\author[inst7,inst8]{Andrea Messina}
\author[inst1]{Cristina	Maria Bernardes	Monteiro}
\author[inst10]{Rafael	Antunes	Nobrega}
\author[inst10]{Igor Fonseca Pains}
\author[inst4]{Emiloano Paoletti}
\author[inst4]{Luciano Passamonti}
\author[inst8]{Sandro Pelosi}
\author[inst9,inst12]{Fabrizio Petrucci}
\author[inst7,inst8]{Stefano Piacentini}
\author[inst4]{Davide Piccolo}
\author[inst4]{Daniele Pierluigi}
\author[inst8]{Davide Pinci}
\author[inst2,inst3]{Atul Prajapati}
\author[inst8]{Francesco Renga}
\author[inst1]{Rita	Joanna da Cruz	Roque}
\author[inst4]{Filippo Rosatelli}
\author[inst4]{Alessandro Russo}
\author[inst1]{Joaquim	Marques Ferreira	dos Santos}
\author[inst4,inst13]{Giovanna Saviano}
\affiliation[inst13]{organization={Dipartimento di Ingegneria Chimica; Materiali e Ambiente; Sapienza Universit\`a di Roma; 00185; Roma; Italy}
}
\author[inst11]{Neil	John Curwen	Spooner}
\author[inst4]{Roberto Tesauro}
\author[inst4]{Sandro Tommasini}
\author[inst2,inst3]{Samuele Torelli}

%\author[inst2]{Author Two}
%\author[inst1,inst2]{Author Three}

%\affiliation[inst2]{organization={Department Two},%Department and Organization
 %           addressline={Address Two}, 
 %           city={City Two},
 %           postcode={22222}, 
 %           state={State Two},
 %           country={Country Two}}

\begin{abstract}
% By measuring not only the energy, but also the direction of the
%nuclear recoils of the atoms of the gas, CYGNO (a CYGNus TPC with Optical readout)
%fits into the wider context of the CYGNUS proto-collaboration, for the development of
%a Galactic Nuclear Recoil Observatory at the ton scale with directional sensitivity.
The CYGNO project aims at the development of a high precision
optical readout gaseous Tima Projection Chamber (TPC) for directional dark matter (DM) searches, to be hosted at Laboratori Nazionali del Gran Sasso (LNGS). CYGNO employs a He:CF$_4$ gas mixture at atmospheric pressure with a Gas Electron Multiplier (GEM) based amplification structure coupled to an optical readout comprised of sCMOS cameras and photomultiplier tubes (PMTs). This experimental setup allows to achieve 3D tracking and background rejection down to O(1) keV energy, to boost sensitivity to low WIMP masses. The characteristics of the optical readout approach in terms of the light yield will be illustrated along with the particle identification properties. The project timeline foresees, in the next 2-3 years, the realisation and installation of a 0.4 m$^3$ TPC in the underground laboratories at LNGS to act as a demonstrator. Finally, the studies of the expected DM sensitivities of the CYGNO demonstrator will be presented.
\end{abstract}




\end{frontmatter}


\section{Directional Dark Matter Search}
\label{sec:sample1}
Since the last decades, dark matter (DM) has been considered a well established element of
our Universe, even though its nature is still elusive and unknown. The leading theory predicts
the existence of at least one new particle not included in the Standard Model of particle physics.
Among various candidates the Weakly Interactive Massive Particles (WIMPs) stand out as they
were predicted by models of both Cosmology and particle physics. Our Galaxy is believed to reside   
within a DM halo made of these hypothetical neutral massive particles which would interact only
weakly with standard matter \cite{SCHNEE_2011}. In this hypothesis, nuclear recoils of few keV can be induced by DM elastic scattering and detected by experiments on Earth. The motion of the Sun around the centre of the Galaxy  produces an apparent wind of DM particles coming from the Cygnus constellation in the laboratory rest frame. This wind imprints a directional dependence in the recoil angular distribution that no background can mimic\cite{Mayet_2016}. The angular distribution will be highly dipolar, an aspect which can be utilised to positively identify DM, to constrain DM halo characteristics and that will help to strongly reduce the impact of the well known neutrino fog on the discovery potential of direct DM experiments \cite{O_Hare_2021}.

\section{The CYGNO detector concept}
The CYGNO experiment aims at building a large volume gaseous Time Projection Chamber (TPC) in a back-to-back configuration with 50 cm drift per side, filled with a He:CF$_4$ 60:40 gas mixture operated at atmospheric pressure and room temperature in the Laboratori Nazionali del Gran Sasso (LNGS) \cite{cygno}. The charge freed by any ionising radiation inside the sensitive volume will be drifted towards the amplification stage which consists in a triple Gas Electron Multiplier (GEM) structure. Here, the charge will be multiplied and, thanks to the properties of CF$_4$, also light will be produced. The readout will be optical by means of two different light detectors:  PMTs and sCMOS cameras by Hamamatsu. The PMT is a fast response detector and will allow to obtain information on the impinging radiation such as the energy, through the amount of photons collected, and the length of the track along the drift direction (henceforth z), thanks to the time spread of the signal. On the other hand, a sCMOS camera is a highly granular sensor with single photon sensitivity which will image the GEM plane, capturing the 2D projection of the track of the original radiation, other than counting the photons for the energy evaluation. Linking the information coming from the two detectors it will be possible to reach a  three dimensional reconstruction of the tracks with a precise measurement of the energy.
%Different prototypes were built in the past to test the characteristics of the amplification stage and the optical readout. With the years the dimension of the readout and especially the drift distance increased reaching the 50 L prototype LIME which already possesses 50 cm drift and is currently being operated underground at the LNGS.
\begin{figure}[!t] 
	\centering
	\includegraphics[width=0.87\linewidth]{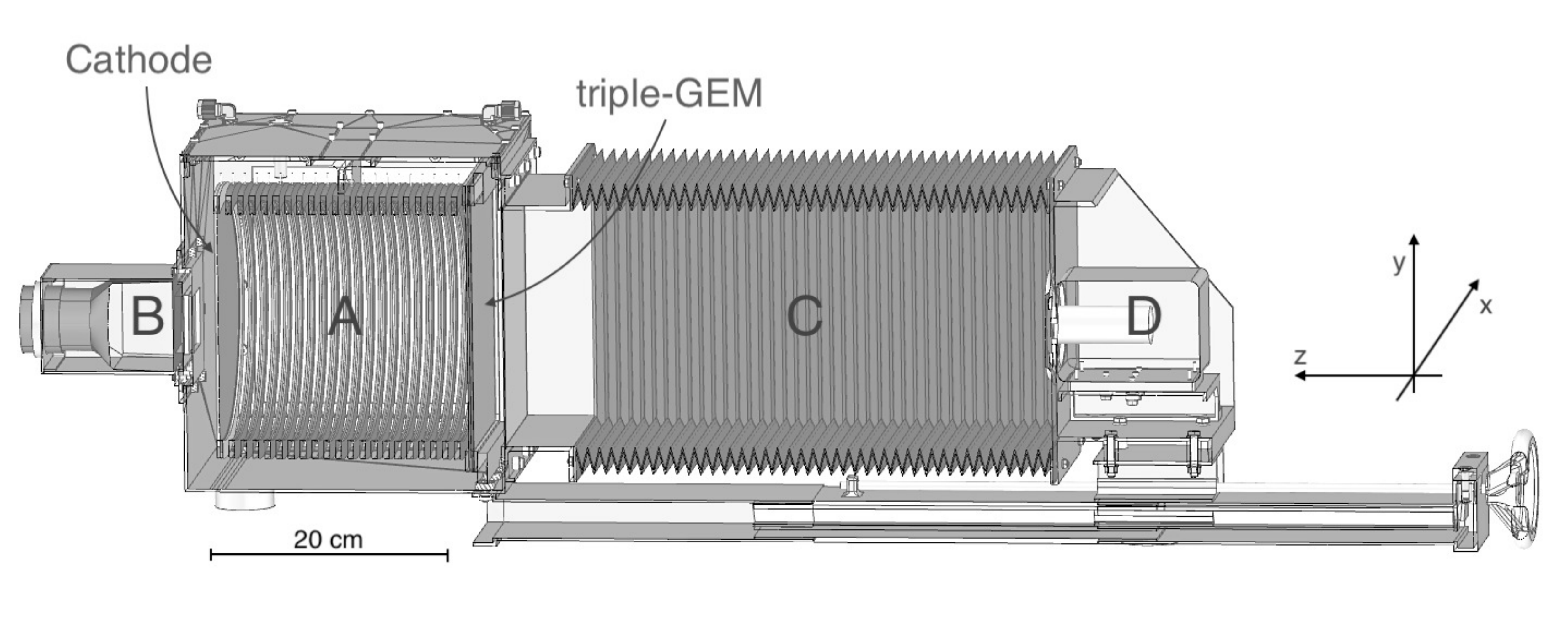}
	\caption{The LEMOn prototype \cite{cygno}. The~elliptical sensitive volume (\textbf{A}), the PMT (\textbf{B}), the optical bellow (\textbf{C}) and the sCMOS camera (\textbf{D}) are indicated.}
	\label{fig:lemonsketch}
\end{figure}
\section{Experimental results from prototypes}
One of the CYGNO prototypes is LEMOn (sketch in Fig. \ref{fig:lemonsketch}) a 20 $\times$ 24 cm$^2$ readout area with 20 cm drift equipped with a triple 50 $\mu$m thick GEM amplification stage. Using an $^{55}$Fe source emitting 5.9 keV photons, it was possible to evaluate a light yield detector response which resulted in roughly 650 photons per keV with an average energy resolution of 15\% along all the drift distances. Such large light yield and the characteristics of the camera permit to infer a 1 keV energy threshold.
%The lower effective threshold was evaluated studying the number of clusters recognized by the reconstruction algorithm \cite{lemon} of the experiment on pictures taken when the sCMOS camera had the sensor cap on. By fitting the exponential behaviour of the number of clusters found as a function of their light intensity, it was possible to find the lower threshold in number of photons that guaranteed that less than 10 clusters per year could arise from noise, an amount considered acceptable for a real data taking. This value is around below 600 photons which correspond to a roughly 1 keV energy threshold.\\
%The capability to discern the absolute distance from the GEM plane in the z direction was estimated by means of a 450 MeV electron beam. These particles were shot at a known distance and parallel to the GEM plane. The granularity of the GEM and sCMOS camera responses allow to image the spatial profile of the tracks and fitting the diffusion the primary electron cloud is subject to along the drift, it was possible to correctly reconstruct the z distance with 15\% resolution.\\
In the context of DM searches, it is of high relevance to discriminate signal nuclear recoils from background electron recoils. LEMOn was exposed to a $^{55}$Fe source which induces electron recoils and to a $^{241}$AmBe neutron source which causes nuclear recoils of few keV. A preliminary study, only exploiting the photon density along the track granted by the granularity of the readout, allowed an efficiency of 18\% on nuclear recoils while suppressing 96\% of background at 6 keV. More thorough studies are ongoing with the help of simulations and neural networks techniques to augment the power of rejection\cite{cygno}.

\begin{figure}[!t] 
	\centering
	\includegraphics[width=1\linewidth]{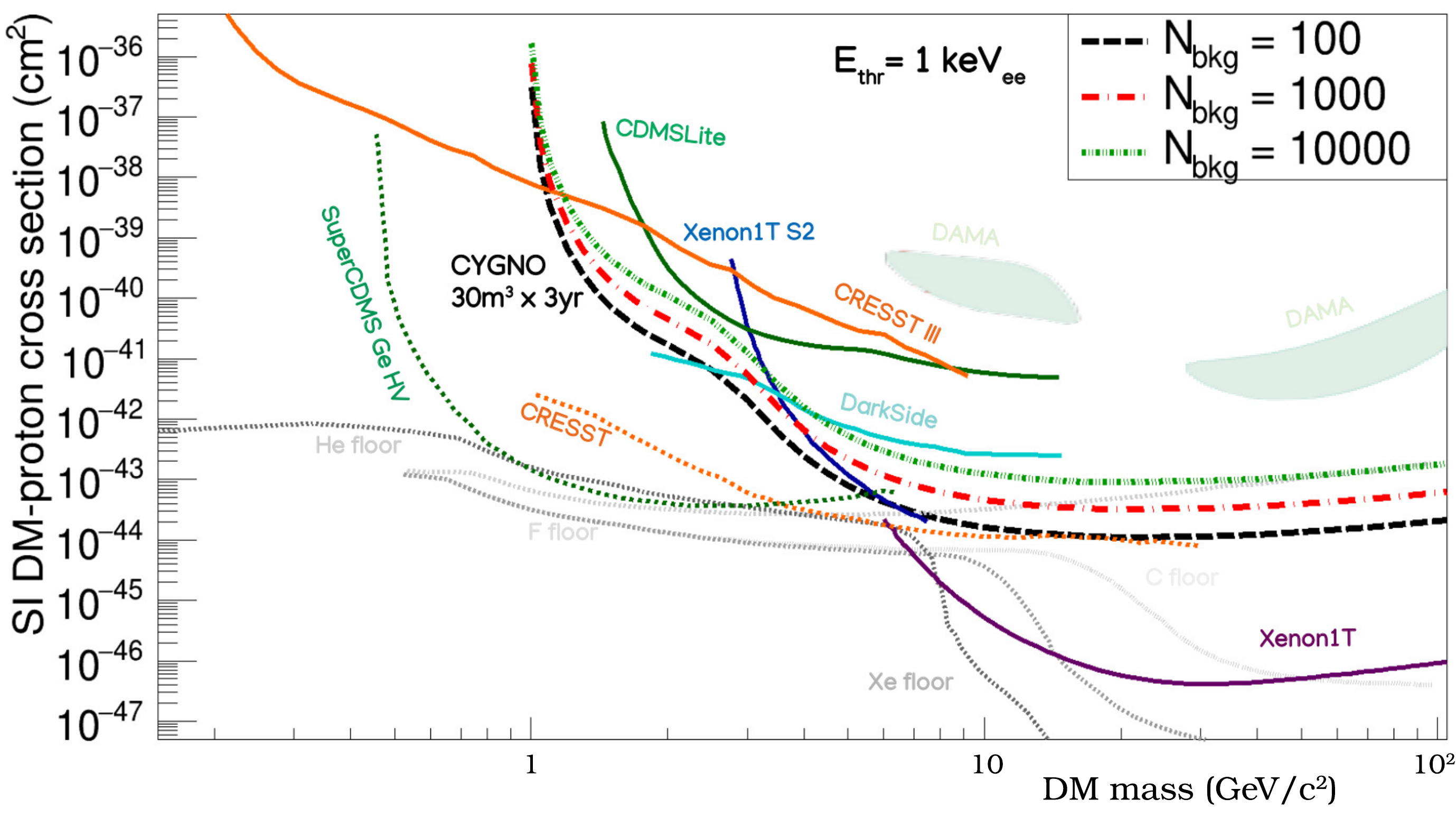}
	\caption{Spin Independent 90\% CI for WIMP-nucleon cross section for 30 m$^3$ CYGNO detector for 3 years of exposure with different background level assumptions and operative threshold of 1 keV$_{\text{ee}}$. Figure from \cite{cygno}.}
	\label{fig:lim}
\end{figure}
\section{Future of CYGNO}
In February 2022, the LIME prototype, a 50 l mono-chamber detector with 50 cm drift length equipped with a triple 50 $\mu$m thick GEM stack, was installed underground at LNGS to be tested under low background environment, realistic for rare event searches. The goal is to validate the Monte Carlo simulations of the expected background with a staged shield configuration of copper (max 10 cm) and water (max 40 cm). In the next future, from 2023 to 2026, CYGNO-04 will be installed underground at LNGS. This detector will comprise a back-to-back configuration with 1 m of drift length split into two 50 cm drift chambers each with a 50 $\times$ 80 cm$^2$ readout area. The technical design report has been submitted to the LNGS and it will be hosted in Hall F. The aim of CYGNO-04 is to prove the scalability of the experimental technique and the capability of enhancing the radiopurity of the materials employed for the construction. Finally, a CYGNO-30 detector, with a back-to-back chamber of 1 m of total drift length with an overall sensitive volume of 30 m$^3$ would be able to sensibly contribute to the DM searches. Fig. \ref{fig:lim} shows the expected limit on the Spin Independent WIMP to nucleon cross section as a function of the DM mass for a CYGNO-30 like detector with a 1 keV$_{\rm{ee}}$ energy threshold and different background considerations, from 100 up to 10$^4$ events per year. The experiment would be competitive with the lowest limits of current experiments below 10 GeV/c$^2$ WIMP masses, but with the uniqueness of being a directional detector.
\section{Ackonwledgements}
Part of this project is funded by the European Union's Horizon 2020 research and innovation programme under the ERC Consolidator Grant Agreement No 818744.

\bibliographystyle{elsarticle-num} 
 \bibliography{biblio}
\end{document}